\documentclass[11pt,a4paper]{article}

\pdfoutput=1
\usepackage{jcappub}
\usepackage{graphicx}
\usepackage{dcolumn}
\usepackage{bm}
\usepackage{color}
\usepackage{calrsfs}
\usepackage{hyperref}
\usepackage{multirow}
\usepackage{float}
\usepackage{soul}  
\input epsf

\newcommand{\be}{\begin{equation}}
\newcommand{\e}{\end{equation}}
\newcommand{\bear}{\begin{eqnarray}}
\newcommand{\ear}{\end{eqnarray}}

\newcommand{\pinocchio}{{\sc PINOCCHIO}}
\newcommand{\pinocchioo}{{\sc PINOCCHIO~}}

\begin{document}

\title{Simulating cosmologies beyond $\Lambda$CDM with \pinocchio}

\author[a]{Luca A. Rizzo,} 
\author[b,c,d]{Francisco Villaescusa-Navarro,}
\author[e,c,d]{Pierluigi Monaco,}
\author[f,e,c]{Emiliano Munari,}
\author[c,d,e]{Stefano Borgani,}
\author[g]{Emanuele Castorina,}
\author[h]{Emiliano Sefusatti}

\affiliation[a]{Institut de Physique Theorique, Universite Paris-Saclay CEA, CNRS, F-91191 Gif-sur-Yvette, Cedex, France}
\affiliation[b]{Center for Computational Astrophysics, 160 5th Ave, New York, NY, 10010, USA}
\affiliation[c]{INAF - Astronomical Observatory of Trieste, via G.B. Tiepolo 11, I-34143 Trieste, Italy}
\affiliation[d]{INFN -- National Institute for Nuclear Physics, Via Valerio 2, I-34127 Trieste, Italy}
\affiliation[e]{Dipartimento di Fisica, Sezione di Astronomia, via G.B. Tiepolo 11, I-34143 Trieste, Italy}
\affiliation[f]{Dark Cosmology Centre, Niels Bohr Institute, University of Copenhagen, Juliane Maries Vej 30, DK-2100 Copenhagen, Denmark}
\affiliation[g]{The Abdus Salam International Center for Theoretical Physics, Strada costiera, 11, I-34151 Trieste � Italy}
\affiliation[h]{INAF, Osservatorio Astronomico di Brera, Via Bianchi 46, I-23807 Merate (LC)  Italy}

\emailAdd{luca.rizzo@cea.fr}
\emailAdd{fvillaescusa@simonsfoundation.org}
\emailAdd{monaco@oats.inaf.it}
\emailAdd{munari@dark-cosmology.dk}
\emailAdd{borgani@oats.inaf.it}
\emailAdd{ecastorina@berkeley.edu}
\emailAdd{emiliano.sefusatti@brera.inaf.it}

\abstract{We present a method that extends the capabilities of the PINpointing Orbit-Crossing Collapsed HIerarchical Objects ({\pinocchio}) code, allowing it to generate accurate dark matter halo mock catalogues in cosmological models where the linear growth factor and the growth rate depend on scale. Such cosmologies comprise, among others, models with massive neutrinos and some classes of modified gravity theories. We validate the code by comparing the halo properties from \pinocchioo against N-body simulations, focusing on cosmologies with massive neutrinos: $\nu\Lambda$CDM. We analyse the halo mass function, halo two-point correlation function, halo power spectrum and the moments of the halo density field, showing that \pinocchioo reproduces the results from simulations with the same level of precision as the original code ($\sim5-10\%$). We demonstrate that the abundance of halos in cosmologies with massless and massive neutrinos from \pinocchioo matches very well the outcome of simulations, and point out that \pinocchioo can reproduce the $\Omega_\nu-\sigma_8$ degeneracy that affects the halo mass function. We show that the clustering properties of the halos from \pinocchioo matches accurately those from simulations both in real and redshift-space, in the latter case up to $k=0.3~h~{\rm Mpc}^{-1}$. We finally point out that the first moments of the halo density field from simulations are precisely reproduced by \pinocchio. We emphasize that the computational time required by \pinocchioo to generate mock halo catalogues is orders of magnitude lower than the one needed for N-body simulations. This makes this tool ideal for applications like covariance matrix studies within the standard $\Lambda$CDM model but also in cosmologies with massive neutrinos or some modified gravity theories.}

\maketitle

\section{Introduction}
\label{sec:introduction}

Here we briefly introduce the cosmological constant problem and the proposed theories aiming at solving it. We point out that a generic prediction of those theories is that the linear growth factor/rate depends on scale, a feature shared by the standard $\Lambda$CDM model in presence of massive neutrinos. 

\subsection{Dark energy}

The discovery of the expansion of the Universe \cite{Riess_98,Perlmutter_99,Schmidt_98} has revolutionized our understanding of cosmology. The standard model of cosmology assumes the existence of a cosmological constant: a fluid with an uniform and constant energy density having an equation of state $P_\Lambda=-\rho_\Lambda$ that acts as a repulsive force and that is responsible for the expansion of the Universe on late times. That model is capable of reproducing extremely well a very diverse set of cosmological observations. However, the value of $\rho_\Lambda$ differs significantly from that obtained from the natural scales of the early Universe, and therefore, reproducing its magnitude requires a huge fine-tuning.

In order to avoid that problem, models with mechanisms that naturally mimic  the properties that a cosmological constant induce on cosmology at late times have been proposed. These models encompass 
new fields and interactions in the Universe, General Relativity modifications, cosmological models with extra dimensions, local deviations from the Copernican principle, and back-reaction effects of the formation of cosmic structures on the overall cosmological background (see \cite{Amendola_2013} for a recent review). For some classes of theories, the observable consequences of gravity modification amount to a variation of the linear growth factor $D(t)$ (and the linear growth rate, its logarithmic derivative with respect to the scale factor $a$, $f=d\log D/d\log a$) that can take a scale dependence: $D(t,k)$.

A way to disentangle different models is through the spatial distribution of galaxies in the Universe. The reason behind is that galaxy clustering is sensitive to the underlying matter distribution and to the growth of matter perturbations, with different models predicting different outcomes. Several missions such as the Dark Energy Survey (DES) \cite{DES}, the Large Synoptic Survey Telescope (LSST) \cite{Ivezic_2008} and EUCLID \cite{Laureijs_2011} will shed light on the nature of dark energy by means of photometric and spectroscopic galaxy surveys.

\subsection{Massive neutrinos}

The presence of a linear scale-dependent growth is not limited to rather exotic models of modified gravity, but it also appears in the standard $\Lambda$CDM model when massive neutrinos are taken into account.  We now provide a brief introduction on the importance of constraining the neutrino masses from cosmological observables, in particular from galaxy clustering.

The standard model of particle physics describes neutrinos as fundamental massless particles organized in three different families. The discovery of the neutrino oscillation phenomenon has however pointed out that at least two of the three neutrino families have mass \cite{Fogli, Tortola} and have established a lower bound on the sum of the neutrino masses: $M_\nu\equiv\sum m_\nu\gtrsim0.06~{\rm eV}$. This has been one of the most important discoveries of the last decades as it points towards the existence of physics beyond the standard model. On the other hand, being neutrinos the second most abundant particles in the Universe (see e.g. \cite{Cosmology_Weinberg}), the fact that they are massive has important consequences for the growth and spatial distribution of matter on cosmological scales. 

Determining the neutrino masses and their hierarchy is thus one of the most important goals in modern physics. Unfortunately, setting tight upper bounds on neutrino masses from laboratory experiments is a great challenge; current bounds on electron neutrino mass, from tritium $\beta$-decay, are $m_{\nu_e}<2.05~{\rm eV}~ (95\%)$ \cite{Kraus_2005, Aseev_2011} and it is expected that in the near future KATRIN\footnote{https://www.katrin.kit.edu/} will lower that limit to $m_{\nu_e}<0.2~{\rm eV}~ (90\%)$. 

On the other hand, neutrino masses leave their signatures in many different cosmological observables. Therefore, it is also possible to constrain the sum of the neutrino masses using data from cosmological observations. In order to achieve this, it is very important to accurately understand, from the theoretical side, the impact that neutrino masses induce on cosmological observables, both at the linear, mildly non-linear and fully non-linear level.

At linear order, massive neutrinos induce two different effects on cosmology \cite{LesgourguesPastor, LesgourguesBook}. On the one hand they modify the matter-radiation equality time, that takes place at lower redshift in a massive neutrino cosmology than in the equivalent model with massless neutrinos. On the other hand, the growth of cold dark matter (CDM) perturbations on small scales is slower in a universe with massive neutrinos. These two effects leave distinct signatures on the anisotropies of the cosmic microwave background (CMB) and observations of those can thus be used to place constraints on the neutrino masses \cite{Planck_2015}. Furthermore, the combination of the above two effects induces a suppression of power, on small scales, in the linear matter power spectrum at low-redshift in a massive neutrino cosmology with respect to the one of the massless neutrino model. Therefore, galaxy clustering
can be used to determine the neutrino masses.

The effect induced by the neutrino masses is therefore embedded, among other probes, in CMB and large-scale structure data \cite{Hannestad_2003, Reid, Thomas, Swanson, Saito_2010, dePutter, Xia2012, WiggleZ, Zhao2012, Costanzi, Basse, Planck_2013, Beutler_2014, Wyman_2013, Battye_2013, riemer, Hamann_2013, Costanzi_2014, Giusarma_2014, Palanque-Delabrouille:2014jca, 2015JCAP...11..011P, Cuesta_2015}. 
The current tightest constraint, $\sum m_\nu<0.12~{\rm eV}$ (95\% C.L.), has been achieved by combining data from the Ly$\alpha$-forest, CMB and baryonic acoustic oscillations (BAO).

The amount of information contained in the linear regime is very large, but the one embedded into the mildly-non linear regime is much larger. Therefore, we can significantly improve our constraints on the value of the cosmological parameters, and in particular on the sum of the neutrino masses, by making accurate theoretical predictions in that regime. Recently, a huge effort has been carried out aiming at investigating, characterizing and finally understanding the impact of neutrino masses on the mildly and fully non-linear regime \cite{LoVerde_2013, Abazajian_2005, Roncarelli_2014, Fontanot_2014, Baldi_2013, Ma, Wong, Brandbyge_haloes, Brandbyge_2008, Paco_2011, Paco_2012, Paco_2013, Castorina_2013, Costanzi_2013, Rossi_2014, LoVerde_2014a, LoVerde_2014b, Ichiki_Takada, Agarwal2011, Bird_2011, Wagner2012, Viel_2010, Villaescusa-Navarro_2012, Marulli_2011, Yacine-Bird, Blas_2014,Peloso_2015,Castorina_2015,Massara_halo_model,Massara_voids}. Most of those studies involve running dedicated N-body simulations incorporating massive neutrinos, which are computationally expensive. Upper limits on the sum of the neutrino masses are expected to further shrink with future surveys as Euclid \cite{Carbone_2011, Costanzi, Audren_2013, Sartoris_2015, Petracca_2015}, DESI \cite{Font-Ribera_2014} or SKA \cite{Pritchard_2008, Villaescusa-Navarro_2015b}. 

\subsection{Purpose}

A key ingredient required to transform galaxy clustering measurements into cosmological parameter values and errorbars is the covariance matrix of measurements, whose inverse enters the likelihood used to compute confidence levels for parameters. Accurately modeling that matrix requires running thousands of N-body simulations, which is computationally prohibitive in most of the cases. In order to circumvent this problem, approximate methods have been developed as {\sc PTHALOS} \cite{PTHALOS}, Augmented Lagrangian Perturbation Theory (ALPT) \cite{Kitaura_2013}, PerturbAtion Theory Catalog generator of Halo and galaxY distributions (PATCHY) \cite{PATCHY}, Comoving Lagrangian Acceleration method (COLA) \cite{Tassev_2013,Tassev_2015,L-PICOLA}, Effective Zel'dovich approximation mocks (EZmocks) \cite{EZmocks}, FastPM \cite{FastPM} and {\pinocchio} \cite{Monaco_2002, Taffoni_2002, Monaco_2002b, Monaco_2013} (see \cite{Chuang_2015} for a comparison among the different methods and N-body simulations).

The above methods, including {\pinocchio}, are mostly designed to work for the standard $\Lambda$CDM model, where the linear growth factor and growth rate are scale-independent. 
The purpose of this work is to develop and validate a method that allows \pinocchioo to generate fast and accurate mock halo catalogues in cosmological models where the linear growth factor and the growth rate depend on scale. Our goal is therefore to provide the community with a tool that can be used, among many other things, to estimate covariance matrices, abundance and clustering properties of dark matter halos and determine amplitudes and errors from cross-correlations of cosmological observables with a high accuracy and predicting power in those models.

As we have seen above, examples of such models are modified gravity theories and the standard $\Lambda$CDM model with massive neutrinos. The method that we will outline in the next section is general, and can be employed in any model that exhibits a linear scale-dependent growth factor/rate. We will however focus our attention on cosmologies with massive neutrinos, and we will compare the results of our method against state-of-the-art N-body simulations with massive neutrinos.

The paper is organized as follows. In Sec. \ref{sec:pinocchio} we quickly describe the {\sc PINOCCHIO} code and depict the method used to extend its capabilities to massive neutrino cosmologies. The set of N-body simulations with massless and massive neutrinos run for this project is described in Sec. \ref{sec:simulations}. The comparison between the abundance and clustering properties of dark matter halos from {\pinocchio} and from the simulations is presented in Sec. \ref{sec:results}. Finally, a summary and the main conclusions of this work are outlined in Sec. \ref{sec:conclusions}.

\section{PINOCCHIO and method implementation}
\label{sec:pinocchio}

In this section we briefly describe the \pinocchioo code and present the method we use
to allow it to generate halo catalogues in cosmologies with a scale-dependent linear growth factor and growth rate, focusing in the particular case of models with massive neutrinos.

{\pinocchio} is a semi-analytic Lagrangian code that starts from a
linear density contrast field,
sampled on a regular cubic grid,
and predicts, for each grid element (or particle), the time at which
this is expected to collapse (to go into orbit crossing). This is done
by smoothing the linear density contrast on many scales and using
ellipsoidal collapse to compute the earliest time at which the
particle is expected to collapse. Collapsed particles are then grouped
into halos with an algorithm that mimics their hierarchical formation.
The original code was presented by \cite{Monaco_2002} and the massively
parallel version in \cite{Monaco_2013}. For this work we use the new version
recently described in \cite{Munari_2015}, where displacements are
computed using second-order Lagrangian Perturbation Theory (2LPT).

One of the assumptions\footnote{Notice that this assumption is justified by the fact that in a massless neutrino cosmology, linear theory predicts that the growth factor is scale-independent.} 
on which the original {\pinocchio} code relies is
that the growth factor
$D(t)$, is scale-independent, i.e. all
perturbations grow at the same relative pace. The growth factor is then
used as the time variable for the evolution of perturbations. In
particular, the ellipsoidal collapse routine returns the growth factor
at which the mass element is predicted to collapse. 

Within the $\Lambda$CDM framework, massive neutrinos can be viewed as a secondary, diffuse, matter component with a large free-streaming velocity \cite{Paco_2012,Massara_halo_model}, in contrast to the dominant CDM which is assumed to have negligible thermal velocities at any redshift. A proper calculation of the evolution and clustering of the massive neutrino fluid, together with its interaction and back-reaction with CDM in the fully non-linear regime would require the use of a Boltzmann plus an N-body code. 

However, it is now well established that while massive neutrinos 
contribute to the expansion rate of the Universe their non-linear clustering is
negligible \cite{Ma,Wong,Paco_2011,Paco_2012,Ichiki_Takada,Paco_2013,Castorina_2013,Costanzi_2013}
for reasonable neutrino masses. One of the consequences of this is that
the halo mass function and halo bias in a cosmology with massive neutrinos
can be described in terms of the CDM plus baryons density field (instead of 
the total matter density field) \cite{Ichiki_Takada,Castorina_2013,Costanzi_2013,Castorina_2015}. 
Based on this finding, it is possible to generate mock halo catalogues in cosmologies with massive neutrinos using \pinocchioo by providing it the proper power spectrum.

In a cosmology with massive neutrinos, the evolution of matter perturbations 
is affected by the neutrino thermal velocities in such a way that the growth factor is both time and scale-dependent: $D(t;R)$. We have adapted our code to deal with a scale-dependent linear growth factor, inherent to massive neutrino cosmologies as well as to some classes of
modified gravity theories.

We now describe in detail the modifications performed to \pinocchio.

(i) Linear growth factors are obtained  from the linear CDM plus baryons power spectrum\footnote{This corresponds to the mass-weighted average of CDM and baryons $P_{\rm cb}(k)=\frac{T_{\rm cb}(k)}{T_{\rm m}(k)}P_{\rm m}(k)$ where $T_{\rm m}(k)$  and $P_{\rm m}(k)$ are the linear matter transfer function and power spectrum, respectively. $T_{\rm cb}(k)=(\Omega_{\rm CDM}T_{\rm CDM}(k)+\Omega_{\rm b}T_{\rm b}(k))/(\Omega_{\rm CDM}+\Omega_{\rm b})$ is the CDM plus baryons transfer function.} $P_{\rm cb}(k,t)$. This is generated using the CAMB code \cite{CAMB} for a given cosmological model. We used $N_{\textrm{CAMB}}=150$ output
times, logarithmically equally spaced in scale factor from $z=99$ to $z=0$.

Scale dependence implies that growth factors for density contrast in Fourier space
differs from that for density and for displacements in configuration space, because these are
obtained from different integrals of the power spectrum and in this case the growth factor
cannot be factored out. So in place of a single $D(t)$ function we have at least three functions of $t$ and $k$, for the power spectrum, the density and the velocity or displacement. The growth factor in Fourier
space is computed by taking ratios of $P_{\rm cb}(k,t)$ at different times: $D^2(t,k)=P_{\rm cb}(k,t)/P_{\rm cb}(k,t_0)$.
Growth factor for density, smoothed at the scale $R$ (the same scales
used to compute collapse times) with a Gaussian
filter $W({\mathbf x};R)$ is
\begin{equation}
D_\delta^2(t,R) = \frac{\int_0^\infty P_{\rm cb}(k,t) \tilde{W}^2(kR) k^2 dk}
{\int_0^\infty P_{\rm cb}(k,t_0) \tilde{W}^2(kR) k^2 dk}\, ,
\label{eq:growdelta}\end{equation}
\noindent
where $\tilde{W}(kR)$ is the Fourier transform of the window function
and the normalization of the growth factor is assumed to be $D_\delta(t_0,R)=1$ for all
$R$, $t_0$ being a reference time, typically corresponding to the last
required output. For displacements:
\begin{equation}
D_v^2(t,R) = \frac{\int_0^\infty P_{\rm cb}(k,t) \tilde{W}^2(kR) dk}
{\int_0^\infty P_{\rm cb}(k,t_0) \tilde{W}^2(kR) dk}\, .
\label{eq:growvel}\end{equation}

(ii) Growth factor is not used as a time variable any more. When the 
ellipsoidal collapse function is called for a smoothing radius $R$, it returns
the growth factor $D_{\rm coll}$ at which the particle is expected to
collapse. Then the collapse redshift is computed by inverting the
$D_\delta(z,R)$ function (equation~\ref{eq:growdelta}) at fixed $R$,
and $1+z_{\rm coll}$ is used in place of the inverse growth factor of
the original code.

(iii) Second-order growth factor of displacements for 2LPT is implemented using the
approximation \citep{Bouchet_1995}
\begin{equation}
D_v^{(2)}(t,R) = \frac{3}{7} D_v^2(t,R) \Omega_m^{-0.007}~.
\label{eq:2lpt} \end{equation}
We notice that the above expression has been
derived for a neutrinoless cosmology. Developing second order Lagrangian perturbation theory
in presence of massive neutrinos is not trivial, so the above formula has to be considered as an 
approximation. We do not however expect large corrections to the above formula for cosmologies with 
realistic neutrino masses, and since we are dealing with second order corrections we consider that
for our purposes here the above formula is accurate enough. We plan to investigate these issues in detail 
in a future work.

(iv) The creation of halo catalogues requires to compute displacements
for particles and groups, both to decide whether accretion or merging
must take place and to compute positions when writing the final
catalog. Displacements are computed using the velocity growth factor of equation~\ref{eq:growvel}.
For collapsing particles, the radius $R$ to be used in the growth factor is the
radius at which the earliest collapse is predicted, for groups we use
their Lagrangian radius ($N_p^{1/3}\Delta $, where $N_p$ is the
number of particles in the halo and $\Delta$ the comoving
inter-particle distance). 

(v) Peculiar velocities are computed using growth rate functions $f(\Omega,R)=d\ln
D_v(a,R)/d\ln a$ and $f^{(2)}(\Omega,R)=d\ln D^{(2)}_v(a,R)/d\ln a$,
obtained numerically from the velocity growth factors.

Calibration of the parameters that regulate the construction of halos has been performed so as to optimize the agreement with the $\Lambda$CDM N-body simulation. This reproduces  the analytic fit of \cite{Crocce_MF} only at the 10\% level, our calibration provides agreement at 5\% level at $z\le1$, so we are in fact fitting the discrepancies of simulations with respect to the analytic formula.\footnote{We achieve this using the following parameters: $f = 0.44$, $e = 0.85$, $s_a = 0.65$, $s_m = 0.12$, $D_0 = 1.4$; we refer to \cite{Munari_2015} for the meaning of parameters and for details on the calibration procedure.}

\begin{figure}
\begin{center}
\includegraphics[width=0.7\textwidth]{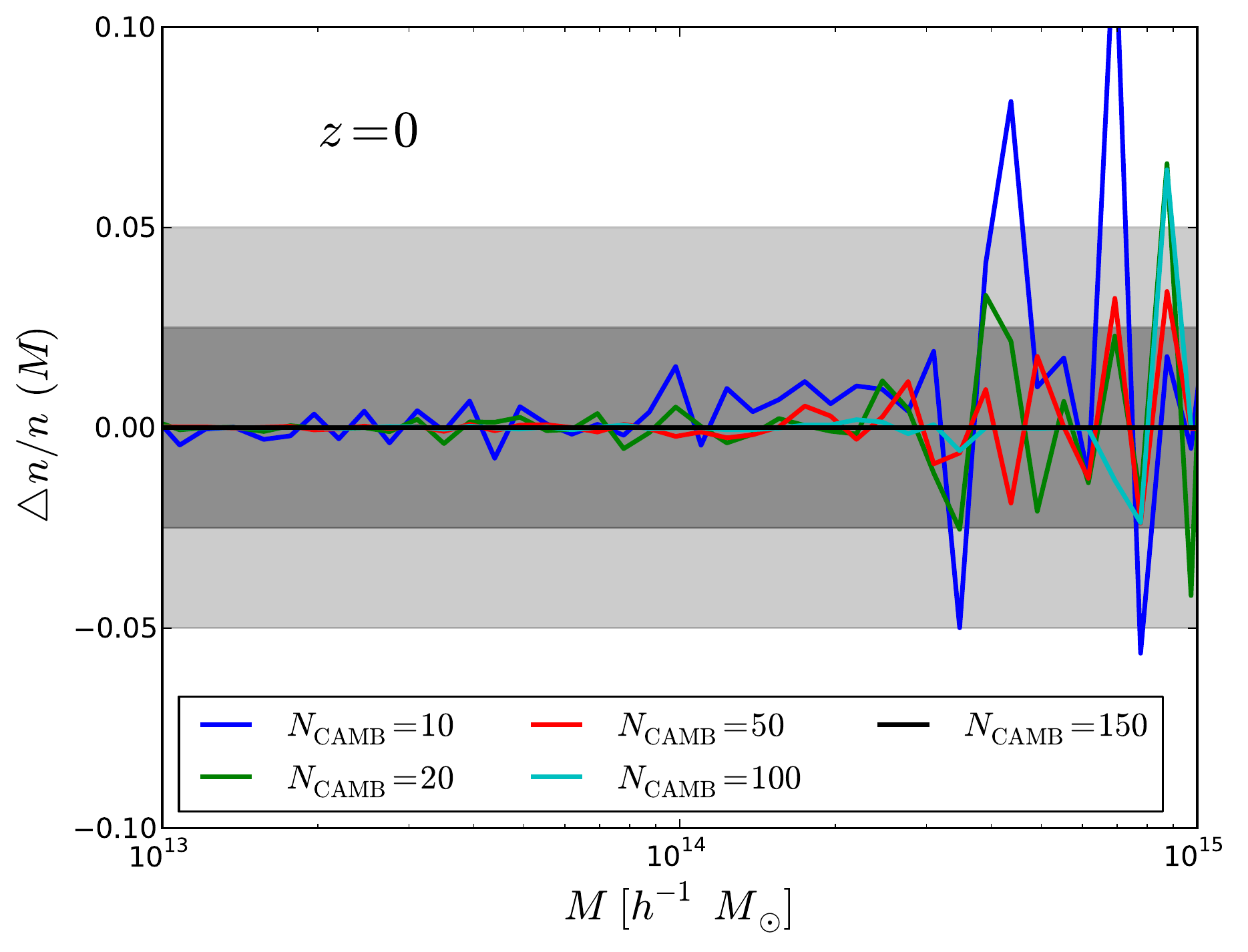}\\
\end{center}
\caption{Impact of the number of CAMB outputs used to compute growth factors on the halo mass function. We show the relative difference in the halo mass function $\Delta n / n (M) = n_{N_{\textrm{CAMB}}}/n_{N_{\textrm{CAMB}}=150}$ - $1$ at $z=0$ for \pinocchioo runs with $M_{\nu} = 0.9$ eV and $N_{\rm CAMB} = 100$ (red), $50$ (blue), $20$ (black) and $10$ (green).} 
\label{fig:GFtest}
\end{figure}

In order to check the robustness and validity of our method we carried out several tests. 
We now briefly discuss the most relevant ones that we performed for a massive neutrino cosmology. 
First, we used the total matter power spectrum in place of the CDM+baryons one as input for \pinocchio. 
Second, we used the density growth factor in place of the velocity one to compute displacements. 
We obtained significantly worse agreement with the halo mass function from N-body simulations 
in both cases. Third, we checked that results are
insensitive to the choice of the reference redshift used to normalize
the growth factors. Fourth, we tested the stability of results against the
number of CAMB outputs, $N_{\textrm{CAMB}}$, used to compute the growing
modes. Figure~\ref{fig:GFtest} shows the relative changes of the mass
function at $z=0$ obtained by using $N_{\textrm{CAMB}}=100$, $50$, $20$ and
$10$, all equally spaced in $\log a$ from $z=99$ to $z=0$. Results are
very stable, to better than 5 \%, down to $N_{\textrm{CAMB}}=20$, while larger
differences are obtained for smaller values. By using $N_{\textrm{CAMB}}=150$
we are thus adopting a very conservative value; converged results at percent level are
achieved with a lower number of CAMB outputs.

\section{N-body simulations}
\label{sec:simulations}

In this section we describe the set on N-body simulations we have run for this paper. We have used these simulations to study how well our extension of {\pinocchio} works for cosmologies with massive neutrinos.

Our simulation suite comprises four different runs, each of them for a different model with different sum of the neutrino masses $M_\nu=\{{\rm 0.0~eV, 0.3~eV, 0.6~eV, 0.9~eV}\}$. Relatively high values are adopted to obtain sizeable differences in the configuration we use.
The simulation box size is 1000 comoving $h^{-1}$Mpc and within it we follow the evolution of $512^3$ cold dark matter particles plus $512^3$ neutrino particles (only for the massive neutrino models). 
When generating the initial conditions, we keep the module of the $k$-space modes equal to their average value given by the power spectrum, so as to keep sample variance low.
The gravitational softening length for each particle type is set to $1/40$ of their mean linear inter-particle distance. We save snapshots at redshifts $z=\{2,1,0.5,0\}$.

The initial conditions are generated at $z=99$ using the Zel'dovich approximation. Linear matter power spectra and transfer functions are computed using the {\sc CAMB} code \cite{CAMB}. The mass-weighted power spectrum from the CDM and baryons power spectra, $P_{\rm cb}(k,z)$, is used to set up the initial conditions of the CDM particles, while for the neutrino particles we employ the neutrino power spectrum. Neutrino particles receive two different velocity components: a peculiar one arising from the gravitational potential and a second one to account for their thermal velocities. For the latter we set the amplitude by randomly sampling Fermi-Dirac momentum distribution of neutrinos, while the direction of the velocity vector is randomly chosen. 

The values of the following cosmological parameters are the same for all runs: $\Omega_{\rm m}=
0.3175$, $\Omega_{\rm b}=0.049$, $\Omega_\Lambda=0.6825$, $h=0.6711$, $n_s=0.9624$, $A_s=2.13\times10^{-9}$ (in agreement with values provided by the Planck collaboration in 2015 \cite{Planck_2015}).
In the model with massless neutrinos $\Omega_\nu=0$ while in cosmologies with massive neutrinos $\Omega_\nu=M_\nu/(94.1h^2~{\rm eV})$. We set the value of $\Omega_{\rm CDM}$ to be $\Omega_{\rm m}-\Omega_\nu-\Omega_{\rm b}$. Since the value of the scalar amplitude is the same in all models, the value of $\sigma_8$ will be different in each cosmology: $\sigma_8=\{0.834, 0.760, 0.690, 0.628\}$ for the models with $M_\nu=\{{\rm 0.0~eV, 0.3~eV, 0.6~eV, 0.9~eV}\}$, respectively. 

Besides the above four simulations we have run an additional simulation for a massless neutrino model. The value of the cosmological parameters of that simulation are exactly the same of the above model with 0.0 eV neutrinos, with the exception of $A_s=1.577\times10^{-9}$ and $\sigma_8=0.717$. We use that simulation to study the well-known degeneracy between $M_\nu$ and $\sigma_8$ in the halo mass function. The value of $\sigma_8$ of this simulation matches the value of $\sigma_8$ computed from the CDM plus baryons linear density field of the simulation with $M_\nu=0.6$ eV. In \cite{Castorina_2013} it was argued that this situation maximizes the degeneracy. 

In all simulations we identify dark matter halos by using the Friends-of-Friends (FoF) algorithm \cite{FoF}, taking $b=0.2$ as the value of the linking length parameter. For our analysis we only consider halos containing at least 32 CDM particles. For cosmologies with massive neutrinos we identify the dark matter halos considering only the CDM field, as the presence of massive neutrinos can be safely neglected \cite{Paco_2012, Castorina_2013, Massara_halo_model}. The masses of the FoF halos are further corrected to take into account the bias induced by particle discreteness according to \cite{Warren_2006}. 

\section{Results}
\label{sec:results}

In this section we present and discuss the main results of our analysis. We compare the properties of dark matter halos from \pinocchioo against those from simulations for different cosmologies with massless and massive neutrinos. The comparison is carried out by running \pinocchioo with the same cosmological models and random seeds as the N-body simulations, again keeping the module of $k$-space modes equal to their average value given by the power spectrum.
In subsection \ref{subsec:MF} we show the abundance of halos as predicted by {\pinocchio} and compare it against the results from N-body simulations and the fitting formula from Crocce et al. \cite{Crocce_MF}. In subsection \ref{subsec:degeneracy} we investigate whether \pinocchioo is able to reproduce the degeneracy between $M_\nu$ and $\sigma_8$ that affects the halo mass function. The clustering properties of the dark matter halos from {\pinocchio} and their comparison with simulations is presented in the subsection \ref{subsec:clustering}.

\subsection{Halo abundance}
\label{subsec:MF}

\begin{figure}
\begin{center}
\includegraphics[width=1.0\textwidth]{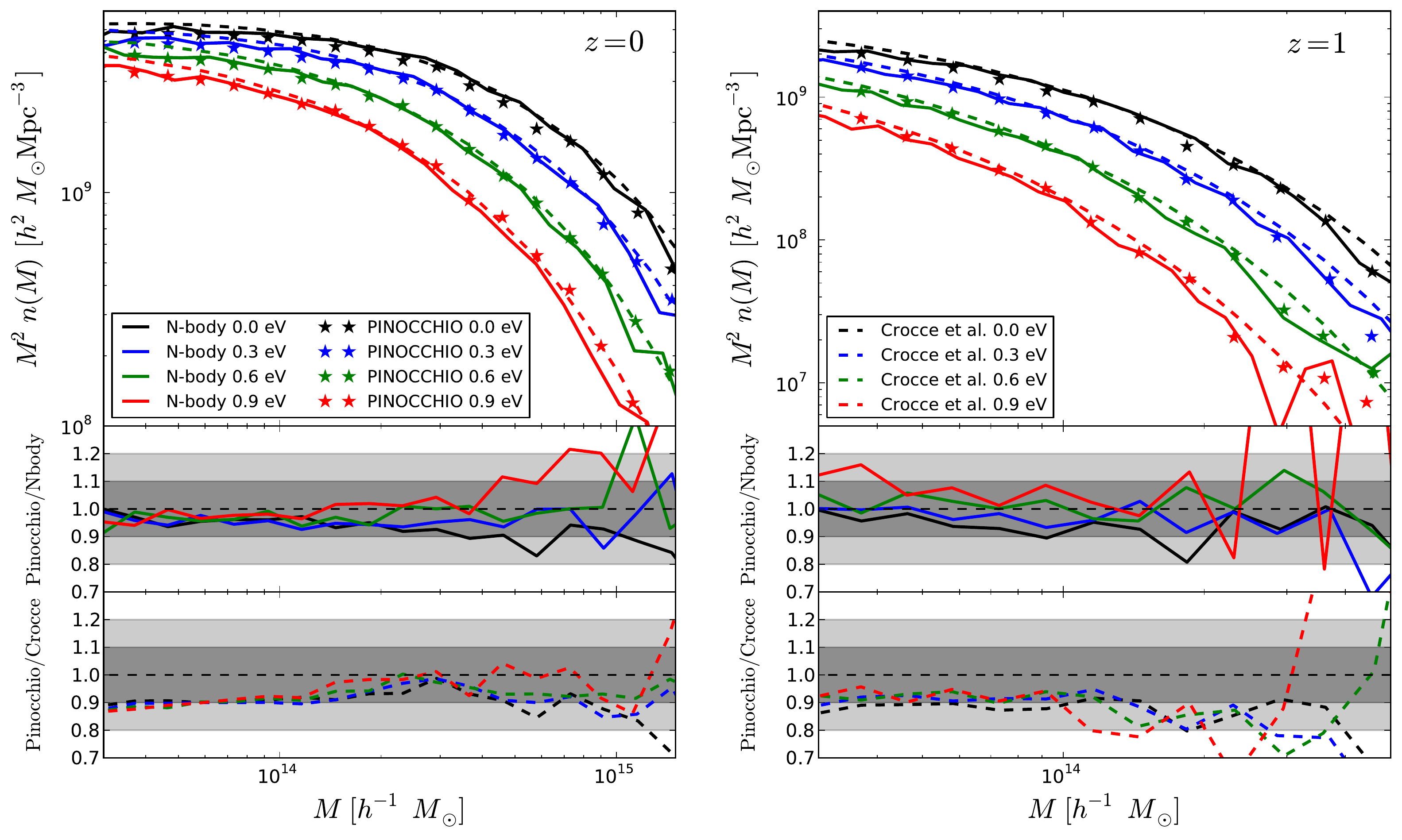}\\
\end{center}
\caption{The upper panels display the comparison between the halo mass function from the N-body simulations (solid lines), the analytic fit of Crocce et al. \cite{Crocce_MF} (dashed lines),  and {\pinocchio} (star points) at $z=0$ (left) and $z=1$ (right) for cosmologies with $M_\nu$ equal to 0.0 eV (black), 0.3 eV (blue), 0.6 eV (green) and 0.9 eV (red). The bottom panels show the ratio between the results from {\pinocchio} and simulations (middle panels) and {\pinocchio} and Crocce et al. (bottom panels).}
\label{fig:MF}
\end{figure}

In Fig. \ref{fig:MF} we show with solid lines the halo mass function, obtained from the N-body simulations for cosmologies with massless and massive neutrinos at $z=0$ (left) and $z=1$ (right). As it can be seen, the higher the neutrino masses the lower the abundance of halos for a fixed mass at any redshift. That behavior is just a consequence of the suppression on the linear matter and CDM+baryons power spectrum induced by massive neutrinos on small scales \cite{Marulli_2011, Villaescusa-Navarro_2012, Ichiki_Takada, Castorina_2013} for models sharing the value of $\Omega_{\rm m}$ and $A_s$. 

In that plot we also display with dashed lines the halo mass function for the different models using the fitting formula of Crocce et al. \cite{Crocce_MF} that was obtained by fitting the abundance of dark matter halos (FoF halos) in the MICE simulations. For cosmologies with massive neutrinos we input the variance of the CDM plus baryons linear density field into the fitting formula instead of the variance of the total matter linear field, since it was pointed out in \cite{Castorina_2013} that this reproduces much better then results of the N-body simulations (see also \cite{Ichiki_Takada, Castorina_2015} \cite{Costanzi_2013}).

It can be seen that the agreement between the analytic fitting formula and the results of the N-body simulations is good at the 10\% level, as already noted in \cite{Castorina_2013}. Differences among the two are due to the low number of halos (in the high-mass end) and to the resolution (in the low-mass end) of the simulations.

The star points in Fig. \ref{fig:MF} show the halo mass function for the different models obtained by employing {\pinocchio}. We find that the agreement between simulations, the analytic fit and {\pinocchio} is very good for all models at the two different redshifts considered in this work. In order to further quantify the agreement between {\pinocchio} and the results from simulations and the analytic fit, we plot in the middle and bottom panel of Fig. \ref{fig:MF} the ratio between those. \pinocchioo has been calibrated to reproduce the mass function of the 0.0 eV simulation to within 5\% in the range of masses ($2\times10^{13}-2\times10^{14}\ h^{-1}M_\odot$) where statistics is good.  Because the agreement between the N-body mass function and the analytic fit is of order 10\%, \pinocchioo is found to underestimates the analytic fit by the same amount.
We find that at $z=0$ the accuracy with which {\pinocchio} reproduces the halo mass function from the simulations remains stable for almost the whole mass range considered here for the four different cosmological models. We obtain deviations larger than $5\%$ between simulations and {\pinocchio} for halo masses larger than $\sim6\times10^{14}~h^{-1}M_\odot$ for the model with $M_\nu=0.9$ eV, where the number of objects that lie in that mass interval is low. Deviations of $\sim10\%$ are also observed in the other models in the high-mass end of the mass function, possibly induced by the low statistics in that regime.

At $z=1$ we find that \pinocchioo is able to reproduce the abundance of dark matter halos from simulations within $5-10\%$ for all cosmological models with a weak dependence on halo mass. Again, the comparison for the $M_\nu=0.9$ eV model is a bit more noisy, in particular in the high-mass end, since the number of halos in that cosmology is much lower than the one from the other models. When comparing the results from \pinocchioo against those from the fitting formula of Crocce et al. at $z=1$ we find again agreement at the $\sim10\%$ level. As with the results at $z=0$, this discrepancy arises due to the imperfect matching between the halo mass function from simulations and the Crocce et al. fitting formula.

Overall we find very good agreement, at the $5-10\%$ level, between the abundance of dark matter halos from N-body simulations, the Crocce et al. analytic fitting formula and {\pinocchio} for all the cosmological models and redshifts considered in this work. In particular, agreement between \pinocchioo and simulations is stable with neutrino mass, showing variations in excess of $\sim10$\% only for the largest and unrealistic neutrino masses.

\subsection{$M_\nu-\sigma_8$ degeneracy}
\label{subsec:degeneracy}

It is well known that a strong degeneracy between the cosmological parameters $M_\nu$ and $\sigma_8$ is imprinted on the halo mass function (see for instance \cite{Carbone_2011,Marulli_2011, Castorina_2013,Castorina_2015},\cite{Costanzi_2013}). This fact severely limits the constraining power of the halo mass function to set, for instance, upper bounds on the sum of the neutrino masses. That degeneracy can however be broken 
by combining CMB and/or BAO observations with the halo mass function \cite{Peloso_2015}.

In this section we investigate how well {\pinocchio} is able to reproduce the mass function of simulations in this degenerate case.
It was pointed out in \cite{Castorina_2013} that cosmologies sharing the value of the cosmological parameters $\{\Omega_m,\Omega_b,\Omega_\Lambda,n_s,\sigma_{8,cb}\}$ will exhibit a very similar, though not identical, halo mass function. $\sigma_{8,cb}^2$ represents the variance of the linear CDM plus baryons power spectrum when smoothed with a top-hat filter of radius $8~h^{-1}$Mpc.

We use two different cosmological models to study this degeneracy: 1) our model with $M_\nu=0.6$ eV massive neutrinos and 2) a massless neutrino model having the same value of the cosmological parameters $\{\Omega_m,\Omega_b,\Omega_\Lambda,n_s,\sigma_{8,cb}\}$ as the former cosmology. We notice that the difference between the latter model and our fiducial massless neutrino cosmology resides in the value of $\sigma_8$, which is 0.717 in the former and 0.834 in the latter.

\begin{figure}
\begin{center}
\includegraphics[width=0.95\textwidth]{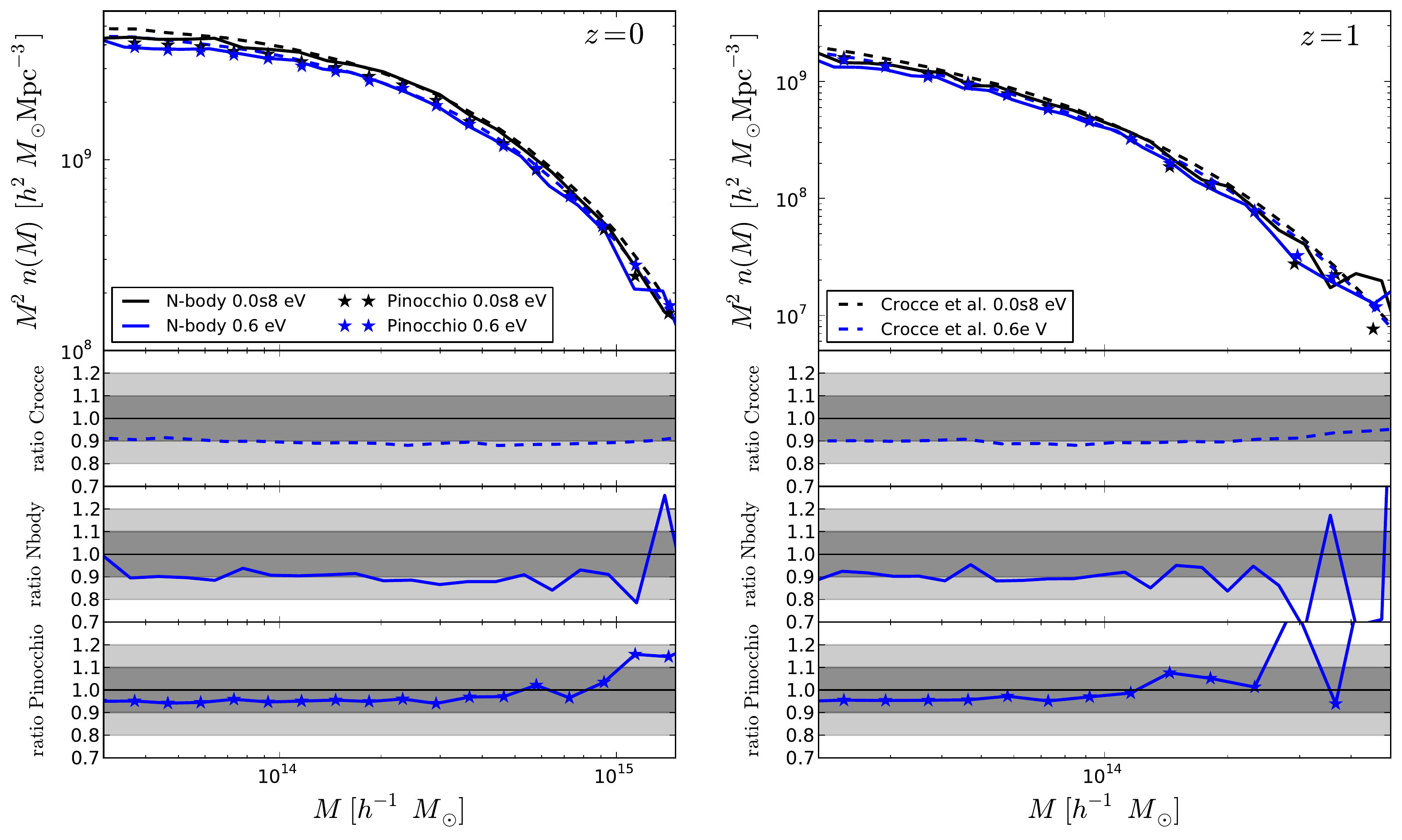}
\end{center}
\caption{Degeneracy between $M_\nu$ and $\sigma_8$ in the halo mass function. We show the halo mass function for a model with $M_\nu=0.6$ eV neutrinos (blue) and a model with massless neutrinos but having a value of $\sigma_{8,cb}$ equal to that of the $M_\nu=0.6$ eV model (black); see text for details. We display results at $z=0$ (left) and $z=1$ (right) for N-body simulations (solid lines), the Crocce et al. fitting formula (dashed lines) and \pinocchioo (stars). The lower panels show the ratio of halo mass function of the two models for Crocce et al. (upper panel), simulations (middle panel) and \pinocchioo (bottom panel).}
\label{fig:degeneracy}
\end{figure}

In Fig. \ref{fig:degeneracy} we show the halo mass function of the two models at redshifts $z=0$ and $z=1$ from \pinocchio, N-body simulations and by using the fitting formula of Crocce et al. It can be seen that the halo mass functions of both models are very similar in shape, while there is a small offset (very hard to observe in practice) in amplitude among the two: the massless neutrino model presents a slightly higher abundance of dark matter halos in comparison with the model with massive neutrinos.

The difference in amplitude can be seen more clearly in the bottom panels of Fig. \ref{fig:degeneracy}, where we show the ratio between the halo mass functions of the models with massive and massless neutrinos. The different panels show the results when the ratio is taken using the outcome of the N-body simulations, \pinocchioo and the Crocce et al. fitting formula. By employing the analytic fitting formula we find that the ratio between the halo mass function of both models exhibits a very weak dependence on halo mass, with a difference in amplitude of $\sim10\%$ at both $z=0$ and $z=1$. These results are corroborated by the ratio of the halo mass function from the N-body simulations, where the same trend and difference in amplitude is observed. Finally, we find that also {\pinocchio} produces two very similar mass functions with a small offset. In this case the offset is found to be of $\sim5$\%, marginally smaller than the one found in simulations. The code is thus able to reproduce both the degeneracy and the small amplitude varitation apart from a 5\% systematic difference when comparing PINOCCHIO and N-body results.

\subsection{Halo clustering}
\label{subsec:clustering}

We now compare the clustering properties of the dark matter halos from the N-body simulations and {\pinocchio}. We focus our attention on the 2-point (hereafter 2pt) statistics, that we measure using two different estimators: the correlation function and the power spectrum. In order to avoid problems related to mass resolution we only consider dark matter halos containing at least 100 particles for both the \pinocchioo and simulation catalogues, corresponding to a mass threshold of $\sim6.6\times10^{13}~h^{-1}M_\odot$. Notice that this value slightly changes among cosmologies since the mass of the CDM particles depends on $\Omega_{\rm CDM}$, which is different in each model.

A thorough test of the ability of {\pinocchio} to reproduce
  clustering of DM halos is presented in \cite{Munari_2015}, where it is
  shown that 3LPT allows to reproduce the halo-halo power spectrum,
  both in real and redshift space, to within 10\% up to
  $k=0.3-0.5~ h~{\rm Mpc}^{-1}$. Conversely halo bias is recovered to within a
  few per cent. In this paper we use 2LPT and a lower mass resolution,
  that was shown in \cite{Munari_2015} to give worse results for
  clustering. As a consequence, 10\% accuracy in $P(k)$ is reached at
  $k=0.2\ h~{\rm Mpc}^{-1}$. Our goal here is to demonstrate that the same
  level of accuracy is reached when using a scale-dependent growth rate for a massive neutrinos cosmology.

\begin{figure}
\begin{center}
\includegraphics[width=1.00\textwidth]{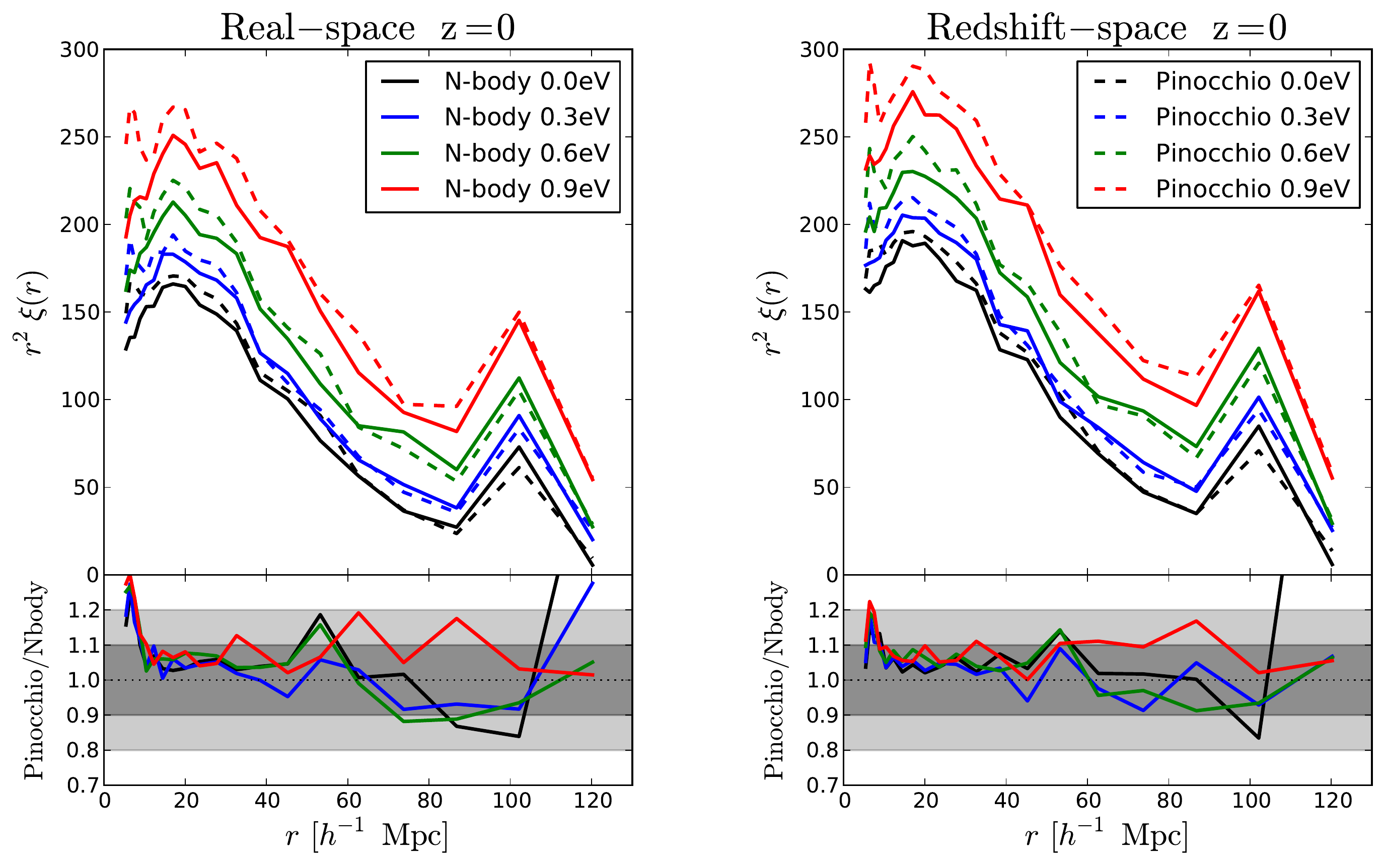}\\
\end{center}
\caption{2-point correlation function of dark matter halos from the N-body simulations (solid lines) and from {\pinocchio} (dashed lines) at $z=0$ in real-space (left) and redshift-space (right) for different cosmological models as shown in the legend. The ratio between the results from {\pinocchio} and from the simulations is shown in the lower part of each panel.}
\label{fig:CF}
\end{figure}

For each cosmological model and for each data set (either from
simulations or {\pinocchio}) we have computed the 2pt correlation
function using the Landy-Szalay estimator \cite{Landy_Szalay} both in
real and redshift-space. The random catalogue we employ for the measurement contains 2 million points, more than 50 times the typical number of halos in the different catalogues. When computing distances among halos we take into account the periodicity of the boxes. In Fig. \ref{fig:CF} we show the results for the different cosmological models in real and redshift-space at $z=0$. We do not show results at $z=1$ since the low number of halos present at that redshift makes our measurements rather noisy.

As shown in Fig.\ref{fig:CF}, at $z=0$ \pinocchioo can reproduce the 2pt correlation function of dark matter halos from simulations within $\sim10\%$ both in real and redshift-space for all the cosmological models considered here. On large-scales results are noisy since the number of pairs relative to the random catalogue decreases at large radii. 
On small scales we appreciate a discrepancy between the clustering amplitude from \pinocchioo and the simulations, amounting to a $\sim2-3$\% of difference in linear bias. This is in line with the uncertainty quoted in \cite{Munari_2015}. This difference is pretty stable with neutrino mass, with the possible exception of $M_\nu=0.9$ eV model where bias gets somewhat larger.

\begin{figure}
\begin{center}
\includegraphics[width=0.95\textwidth]{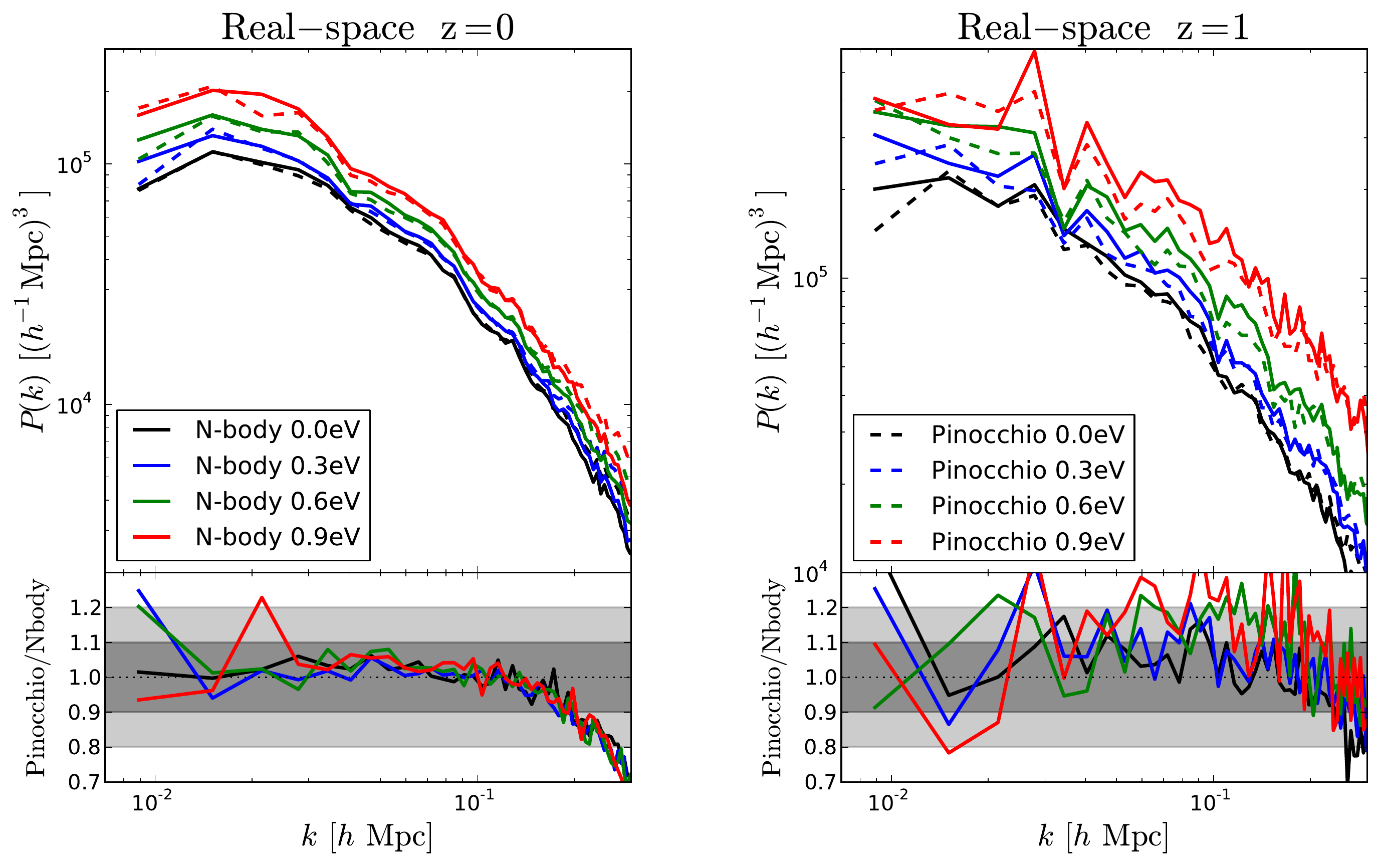}\\
\includegraphics[width=0.95\textwidth]{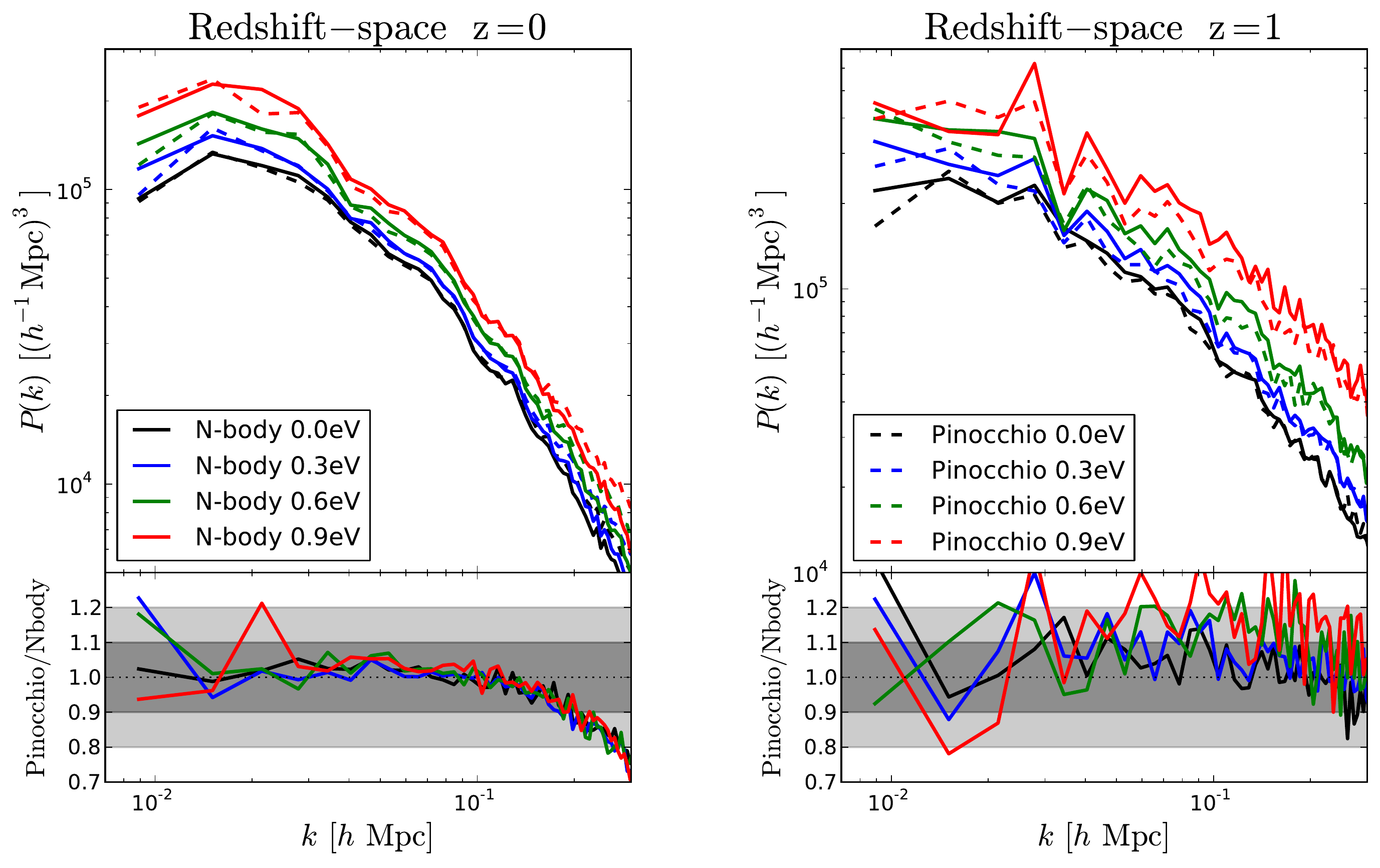}\\
\end{center}
\caption{Power spectrum of dark matter halos from the N-body simulations (solid lines) and from {\pinocchio} (dashed lines) at $z=0$ (left) and $z=1$ (right) in real-space (upper panels) and redshift-space (bottom panels)  for different cosmological models as shown in the legend. In all cases we have subtracted the shot-noise level of the measured power spectra. The ratio between the results from {\pinocchio} and from the simulations is shown in the lower part of each panel.}
\label{fig:Pk}
\end{figure}

We also compare the clustering properties of the dark matter halos from the simulations and {\pinocchio} in Fourier-space by computing the halo power spectrum. In Fig. \ref{fig:Pk} we show the halo power spectrum at $z=0$ and $z=1$ from the N-body simulation and {\pinocchio} catalogues in real and redshift-space for the four different cosmological models with massive and massless neutrinos. We have subtracted the shot-noise level from the power spectrum measurements in all cases.

At $z=0$ we find an excellent agreement, for all cosmological models, between the halo power spectrum from \pinocchioo and the simulations. The agreement, better than $5\%$ for a wide range of scales, breaks down on small scales, for the reasons mentioned above. The halo power spectrum from the \pinocchioo catalogues agree at the $10\%, 20\%$ level on scales as non-linear as $k=0.2, 0.3~h~{\rm Mpc}^{-1}$ both in real and redshift-space.

At $z=1$ the agreement is a bit worse, but still at the level of $10\%$ down to $k=0.3~h~{\rm Mpc}^{-1}$ in real and redshift-space. Consistently with what noticed above, the most pronounced discrepancies between the halo power spectrum from \pinocchioo and the simulations take place for the $M_\nu=0.9$ eV model, where linear bias is overestimated by $\sim10$\%. We notice from Figure~\ref{fig:MF} that the match in the mass functions is less good for the highest neutrino mass, so a fixed cut in mass selects different halo populations that have different bias.

Overall, we find that, for all cosmological models but the one
  with the largest (and unrealistic) neutrino mass, the ability with
  which the {\pinocchio} code recovers the clustering properties of
  the dark matter halos is stable and independent of neutrino mass.
  Small (10\%) discrepancies are found only for the $M_\nu=0.9$ eV
  model.

\subsection{Halo field}
\label{spatial-halo}

\begin{figure}
\begin{center}
\includegraphics[width=0.6\textwidth]{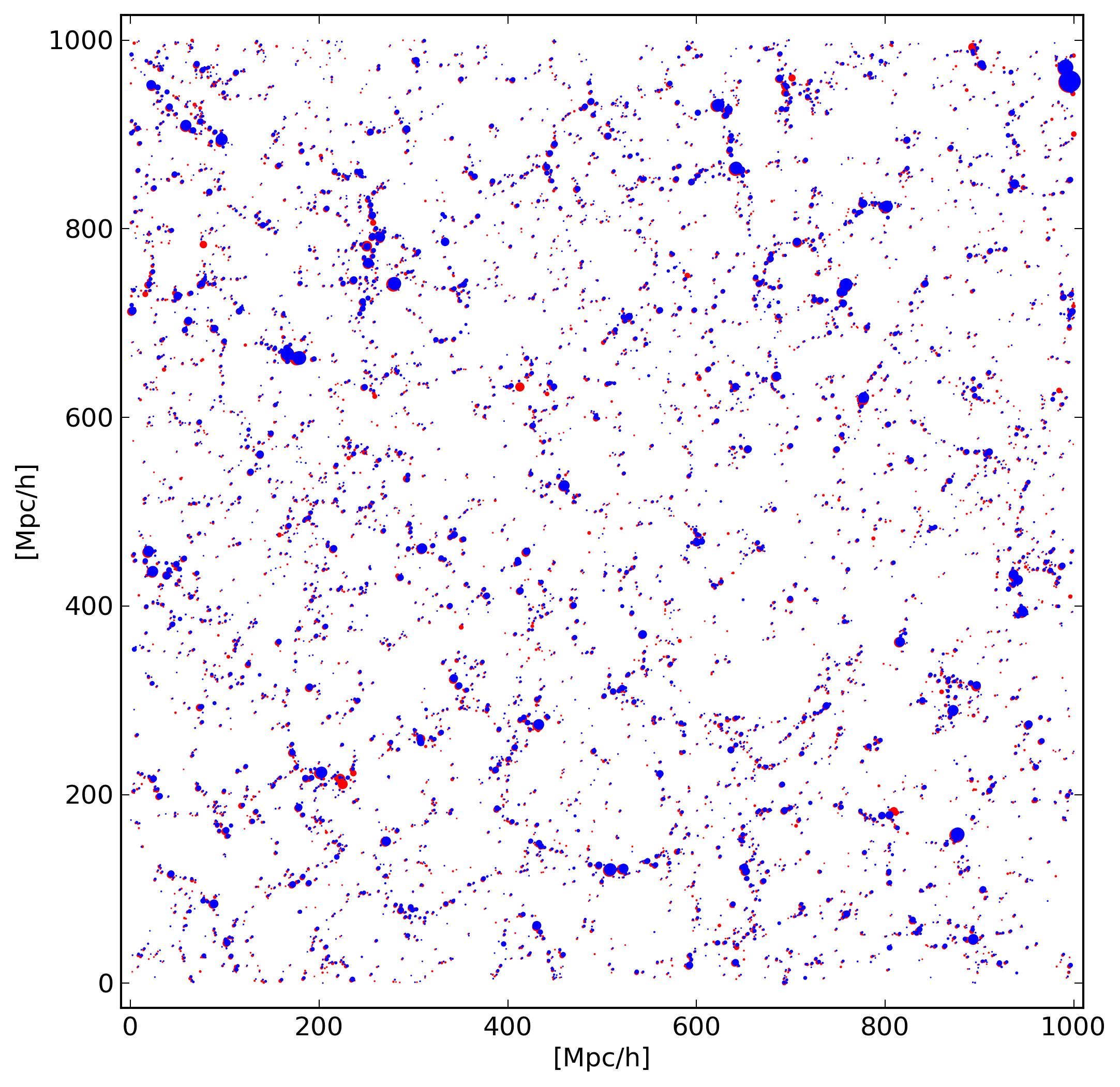}\\
\includegraphics[width=0.6\textwidth]{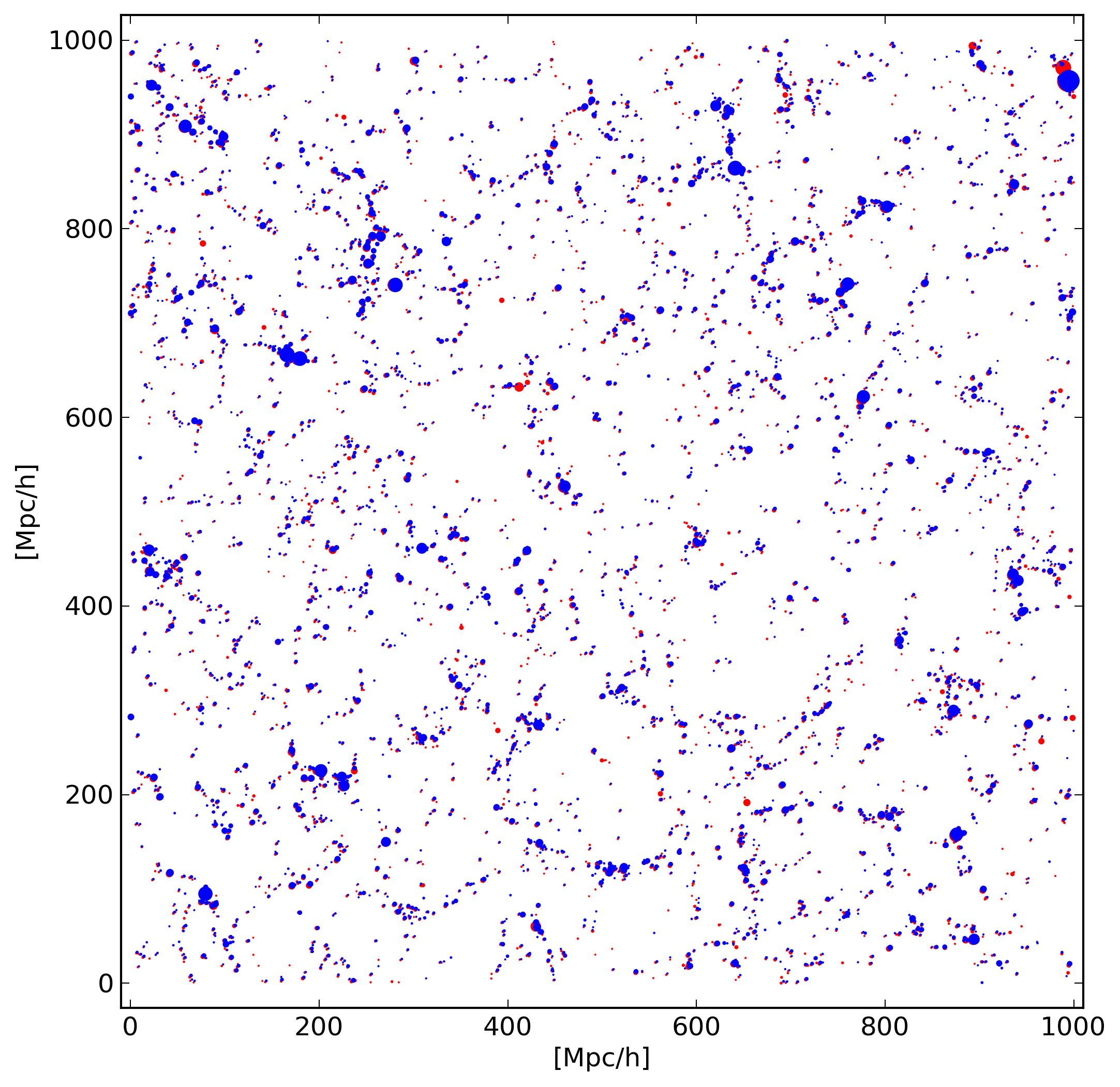}
\end{center}
\caption{Spatial distribution of dark matter halos from {\sc Pinocchio} (blue points) and N-body (red points) for the model with massless neutrinos (top) and $M_\nu=0.9$ eV (bottom) at $z=0$, in a slice of $20~h^{-1}{\rm Mpc}$. In the picture are shown only the halos with masses $ M > 1.8\times10^{13}~h^{-1}M_{\odot }$ for the  massless neutrinos case and with masses $ M > 1.7\times10^{13}~h^{-1}M_{\odot }$ for $M_\nu=0.9$ eV. The size of the points is proportional to the mass of the halo it represents, although the actual size does not coincide with its virial radius.}
\label{fig:halo_comparison}
\end{figure}

In this section we want to test the capability of \pinocchioo of reproducing the properties of the halo field from simulations such as its probability distribution function (pdf). We notice that a detailed test was presented in \cite{Monaco_2013}, for massless neutrino cosmologies, where it is shown that \pinocchioo is able to reproduce well the large-scale structure as compared to simulation, even if most massive haloes in PINOCCHIO tend to be more isolated than their simulation counterparts. Here we want to demonstrate that we can reach the same level of accuracy using a scale-dependent growth rate for a massive neutrinos cosmology: a first visual impression can be given using Fig.\ref{fig:halo_comparison} where compute the halo spatial distribution obtained from \pinocchioo against the one from the N-body simulation for the models with $M_\nu=0.0, 0.9$ eV at $z=0$, in a region of size $1000 \times 1000 \times 20 , (h^{-1}{\rm Mpc})^3$. As it can be seen the agreement between the location and mass of dark matter halos from \pinocchioo and the simulations is remarkable.

\begin{figure}
\begin{center}
\includegraphics[width=0.9\textwidth]{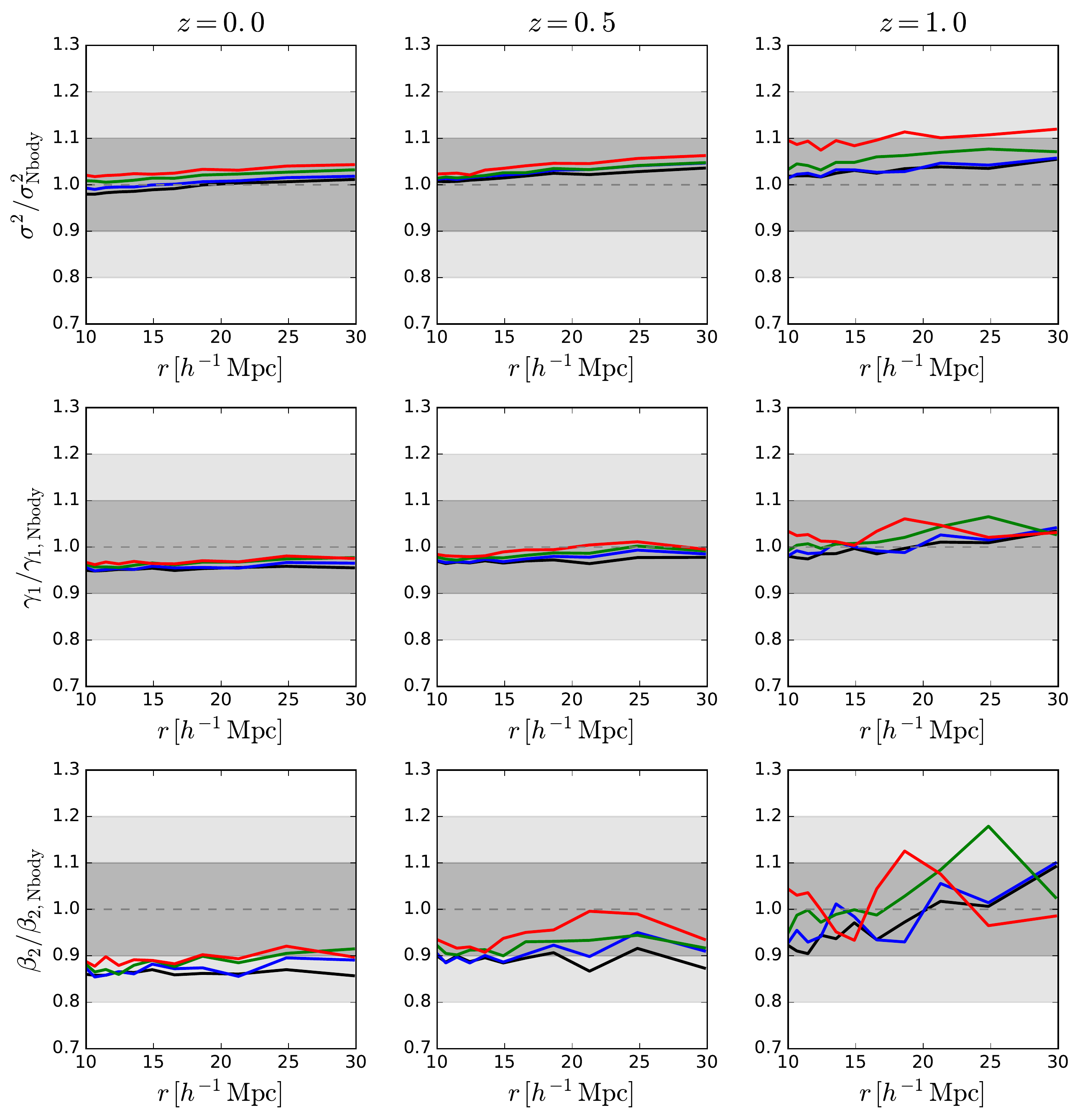}\\
\end{center}
\caption{Moments of the probability distribution function of the halo field as a function of the smoothing scale: the variance (top row), skewness (middle row) and kurtosis (bottom row) at $z=0$ (left column), $z=0.5$ (middle column) and $z=1$ (right column). Each panel shows the ratio between the results from {\sc Pinocchio} and the simulations for $M_\nu=0.0, 0.3, 0.6, 0.9$ in black, blue, green and red, respectively.}
\label{fig:pdf_density}
\end{figure}

In order to test the stability of the last result with neutrino mass in a more quantitative way, we have computed the pdf of the halo density field and its moments: in Fig.\ref{fig:pdf_density}  we show the ratio among the variance $\sigma^2$, the skewness $\gamma_1$ and kurtosis $\beta_2$ computed from the halo catalogues from Pinocchio and the ones obtained from the simulations, for the different cosmological models explored in the paper at $z=0,0.5,1.0$ :  the agreement between the results is stable with neutrino mass, with the exception of the unrealistic $0.9 \, eV$ case where the assumption of treating neutrinos at the linear level becomes weak.

These results show that \pinocchioo is able to reproduce the halo spatial distribution at a remarkable level and these results are stable increasing the neutrino mass, up to the level in which neutrinos cannot be described under the linear assumption. 

\section{Summary and conclusions}
\label{sec:conclusions}

The discovery that neutrinos are massive particles has profound implications in both particle physics and cosmology. From the particle physics side, the fact that neutrinos are massive requires an extension of the standard model of particle physics and to identify the field which gives then mass. For cosmology, it is expected that the growth and spatial distribution of matter in the Universe will be affected by neutrinos, since those are the second most abundant particles in the whole Universe.

One of the most important questions in physics is: what are the neutrino masses? The answer to this question is embedded into cosmological observables such as the spatial distribution of galaxies. Constraints on the value of the cosmological parameters, and in particular on the sum of the neutrino masses, via galaxy clustering, requires to compute the likelihood of a given model given the observed data. One of the fundamental pieces needed to estimate the likelihood is the covariance matrix of clustering statistics, whose accurate evaluation involves running thousands of N-body simulations, which is computationally prohibitive in most cases. In order to circumvent this problem, fast and accurate numerical tools have been developed with the goal of providing catalogues of dark matter halos whose abundance and clustering properties reproducing as close as possible those from N-body simulations. One of those codes is \pinocchioo \cite{Monaco_2002, Taffoni_2002, Monaco_2002b, Monaco_2013}.

The purpose of the present work has been to extend the capabilities of \pinocchioo to be able to generate halo catalogues in cosmologies with a linear scale-dependent growth factor/rate, as models with massive neutrinos or some modified gravity theories. We have achieved this by proposing a new method where growth factors and growth rates are computed from the output of CAMB. In this paper we have focused on cosmologies with massive neutrinos for illustrative purposes. Our method relies on the fact that the abundance and clustering properties of dark matter halos in massive neutrino cosmologies depend only on the properties of the linear CDM plus baryons density field \cite{Ichiki_Takada, Paco_2013, Castorina_2013, Costanzi_2013, Castorina_2015}. The estimated growth factors from CAMB outputs are then used in \pinocchioo  to generate halo catalogues. We have carried out several tests to check the robustness of our method, showing that if \pinocchioo is input with incorrect power spectra or the same growth factor is used for both velocities and density, the properties of halos from \pinocchioo are significantly different to those from simulations.

In order to further validate the code we have compared, for different cosmologies with different neutrino masses, the properties of the halo catalogues generated by \pinocchioo against N-body simulations. We have focused our analysis on the abundance of dark matter halos (halo mass function) and on the clustering properties of those (2pt correlation function and power spectrum). 

After a fine calibration performed on the $M_\nu=0.0$ eV model (aimed at reproducing the discrepancies of the simulation with respect to the analytic fit of \cite{Crocce_MF}), 
we find that \pinocchioo is able to reproduce the halo mass function from N-body simulations within $5\%$ for cosmologies with neutrino masses ranging from $M_\nu=0.3$ eV to $M_\nu=0.9$ eV at both $z=0$ and $z=1$,  with some ($\sim10$\%) difference only for the largest and unrealistic neutrino mass. 

We have also verified that \pinocchioo reproduces the well-known degeneracy between the cosmological parameters $M_\nu$ and $\sigma_8$ that affects the halo mass function. We emphasize that \pinocchioo is able to reproduce not only the absolute amplitude of the mass functions but also the relative differences among halo abundances from different models (to within $5\%$), which, for this particular case are very small.
In both cases we find that \pinocchioo is capable of reproduce the spatial distribution of halos in real-space within $\sim10\%$ at $z=0$ and $z=1$ down to $k=0.2~h~{\rm Mpc}^{-1}$. We have also checked that the clustering properties of dark matter halos from N-body simulations are reproduced, for all the cosmological models considered in this work, at the same accuracy level demonstrated in \cite{Munari_2015}. We have checked this using two different estimators for the halo clustering: the 2pt correlation function and the power spectrum. Finally in order to test the capability of \pinocchioo of reproducing the halo spatial distribution with neutrino masses against simulations, we have computed the moments of the halo density probability distribution function, showing that we can reach the same level of accuracy of \cite{Monaco_2013} with realistic values of the neutrino mass. Furthermore, the peculiar velocities of dark matter halos from simulations are also very well reproduced by \pinocchio, as the clustering properties of dark matter halos from simulations and \pinocchioo agree at the $\sim10\%$ level down to $k=0.2~h~{\rm Mpc}^{-1}$ in redshift-space. We notice that in the case of the power spectrum at $z=0$, \pinocchioo is able to reproduce the clustering of halos from simulations within $5\%$ up to $k\sim0.15~h~{\rm Mpc}^{-1}$ both in real and redshift-space. This is obtained using 2LPT and a relatively modest mass resolution. \cite{Munari_2015} showed that even better results are obtained with 3LPT and working with a better resolution (after which convergence is reached). So, the main result of this paper is not the accuracy level reached but its constancy for $M_\nu\le0.6$ eV, with some $\sim10$\% differences found only for the highest neutrino mass.

The work presented in this paper extends the capabilities of \pinocchio, allowing it to predict the abundance, spatial distribution and peculiar velocities of dark matter halos in cosmologies with a scale-dependent growth rate of linear perturbations like the case of massive neutrinos.  This same method can be applied to those classes of modified gravity theories where the main effect of modification is a specific, scale-dependent growth factor/rate. We emphasize that the time required to obtain a halo catalogue with \pinocchioo is a factor of $\sim1/2000$ lower than the time needed to obtain it from an N-body simulation, thus further supporting its use for an accurate prediction of cosmic covariances of clustering measurements also in non-$\Lambda$-CDM cosmologies.

\section*{Acknowledgements}
N-body simulations have been run in the Zefiro cluster (Pisa, Italy). FVN has been supported by the Simons Foundation and by the ERC Starting Grant ``cosmoIGM''. PM acknowledges support from a FRA 2015 grant of Trieste University. SB acknowledges support from the PRIN-MIUR 201278X4FL grant “and from the PRIN-INAF/2012 Grant “The Universe in a Box: Multi-scale Simulations of Cosmic Structures”. We acknowledge partial support from ``Consorzio per la Fisica -- Trieste'' and from the INFN IS PD51 “INDARK.

\bibliographystyle{JHEP}
\bibliography{Bibliography} 

\end{document}